\documentclass[conference]{IEEEtran}
\usepackage[pdftex]{graphicx}
\usepackage{subcaption}
\usepackage{booktabs}

\usepackage[hidelinks]{hyperref}
\usepackage[utf8x]{inputenc}
\usepackage{amsmath}
\usepackage{amssymb}
\usepackage{textcomp}
\usepackage[usenames,dvipsnames]{color}
\usepackage{balance}

\begin{document}

\IEEEoverridecommandlockouts%

\IEEEpubid{\makebox[\columnwidth]{\color{red}Author pre-print. Paper submitted to 14th IEEE eScience.} \hspace{\columnsep}\makebox[\columnwidth]{ }}

\title{A Scalable Machine Learning System for Pre-Season Agriculture Yield Forecast}
\author{\IEEEauthorblockN{Igor Oliveira, Renato L. F. Cunha, Bruno Silva, Marco A. S. Netto}
\IEEEauthorblockA{IBM Research}
}
\maketitle
\begin{abstract}
Yield forecast is essential to agriculture stakeholders and can be obtained
with the use of machine learning models and data coming from multiple sources.
Most solutions for yield forecast rely on NDVI (Normalized Difference
Vegetation Index) data which, besides being time-consuming to acquire and
process, only allows forecasting once crop season has already started.
To bring scalability for yield forecast, in the present paper we describe a
system that incorporates satellite-derived precipitation and soil properties
datasets, seasonal climate forecasting data from physical models and other
sources to produce a pre-season prediction of soybean/maize yield---with no need of
NDVI data. This system provides significantly useful results by the exempting
the need for high-resolution remote-sensing data and allowing farmers to
prepare for adverse climate influence on the crop cycle. In our studies, we
forecast the soybean and maize yields for Brazil and USA, which corresponded to
44\% of the world's grain production in 2016. Results show the error metrics for
soybean and maize yield forecasts are comparable to similar systems that only
provide yield forecast information in the first weeks to months of
the crop cycle.

\end{abstract}

\IEEEpeerreviewmaketitle%

\section{Introduction}\label{sec:intro}

Agriculture is an industry sector that is benefiting strongly from the
development of sensor technology, data science, and machine learning (ML)
techniques in the latest years. These developments come to meet environmental
and population pressures faced by our society, where reports indicate a need of
strong global agriculture yield increase to provide food for a growing population
in a warmer planet~\cite{fao2016wfp}.

Yield forecast is one of the tasks that can be performed by current ML
algorithms~\cite{dahikar2014agricultural,ji2007artificial,wang2018deep}. Field
sensors, satellites, unmanned aerial vehicles (UAVs), and farming equipment can
provide a significant amount of data on soil conditions, plant physiology,
weather, climate, and several of the processes taking place in a farm. These
datasets allow the creation of classification and forecast models that can be
extremely helpful to agriculture production.

Most of the work done in the field of yield forecasting via ML makes use of some
sort of remote sensing data over the farm, specially in the form of Normalized
Difference Vegetation Index (NDVI)~\cite{van1993relationship}, a popular
indicator of vegetation activity that can be retrieved from near-infrared and
red spectral channels. These indexes have the advantage of providing direct
observations of a farm and can be useful to follow the crop cycle. While these
datasets provide insights on near real time to problems such as diseases and
deficiencies, they allow yield forecasting only after planting occurs, as one
can analyze crop development and try to predict its final outcome after
harvesting.

In this paper, we introduce an ML-based system using data from multiple sources
to perform soybean yield forecasting before the start of the crop
season---process known as \emph{pre-season forecasting}. The system is composed
of a recurrent neural network (RNN) trained with precipitation, temperature,
and soil properties as features and historical observed soybean and/or maize
yield at  municipality level for 1500+ cities in Brazil and USA as labels. We
considered  Brazil and the USA in our case studies as they are two of the
largest crop producers of the world, accounting for 28\% and 35\% of soybean global
production and 6\% and 36\% of maize global production respectively as of
2016~\cite{faostat2016}. Operationally, the meteorological data is provided by
a reanalysis-bases seasonal forecast product of temperature and precipitation,
which allows for forecasting up to seven months in the future. Results are
comparable and in some cases superior to similar models that need remote sensing
data over the farm, thus only capable of providing a forecast in the first
weeks/months of the crop cycle (\emph{early season forecast}).

\smallskip
The two major contributions of this work are:

\begin{itemize}
 \setlength\itemsep{0.5em}

\item A yield forecast system based on fewer data requirements compared to
 existing yield forecast solutions which demand large amounts of remote sensing
 data.  Our system retrieves the necessary climate and soil properties data for
 a given coordinate automatically from an appropriate source. Another advantage is
 that the system works also on large regions, and provides forecasts at a resolution
 compatible with best input data resolution, which in the case is 250m
 originally from the soil data.

\item The capability of forecasting yield before the beginning of the crop season.
This provides users the capability to perform strategy changes, like choosing a
more robust genetic variation before planting or even changing the crop type, in order to
accommodate for extreme climatic variations further ahead in the crop cycle.

\end{itemize}

\begin{figure*}[!h]
  \includegraphics[width=.7\linewidth]{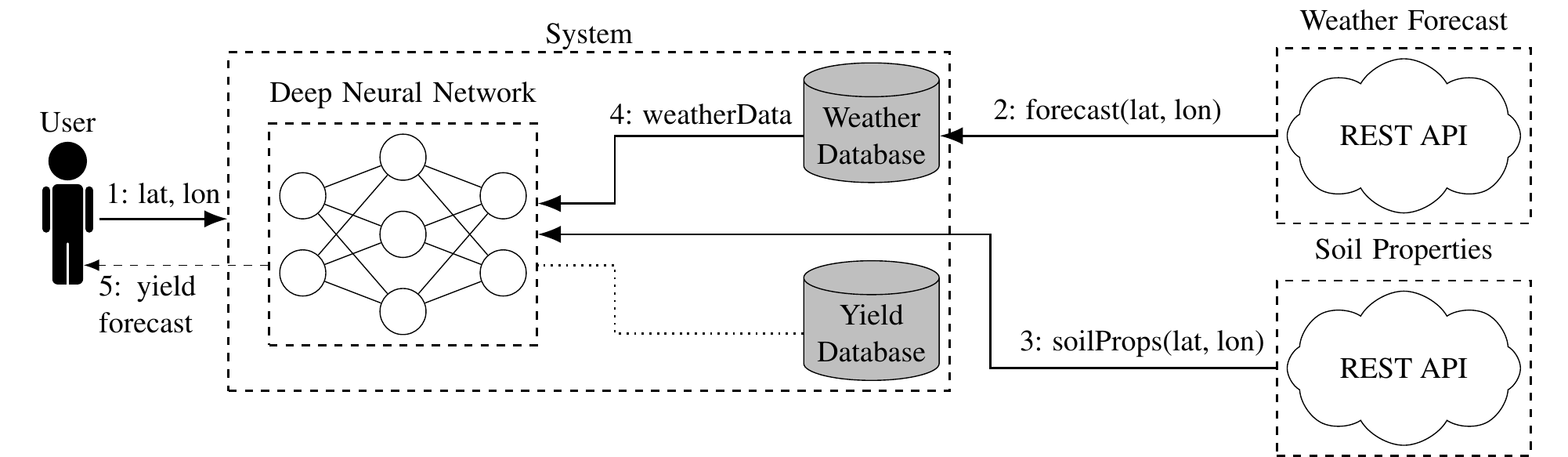}
  \centering
  \caption{%
      System Architecture. The system we built can perform yield forecasts on
      the fly. It is able to do so by, upon user request (1), downloading
      cached weather forecast (2, 4) and soil properties data from RESTful APIs
      (3), and combining them to issue yield forecasts (5) using a trained deep
      neural network model.  Both the APIs and the system work with bounding
      boxes for issuing yield forecasts for \emph{regions} over the globe.
  }\label{fig:system}
\end{figure*}

The scalability for yield forecast comes from our observation that
a neural network model can detect and exploit redundant information coming from soil and weather data.
We also observed the neural network model can learn an implicit representation of the cycles of the crops, which is further detailed in this paper.

The paper is organized as follows. In Section~\ref{sec:related}, we discuss
existing work on yield forecast. We introduce the proposed system containing
the model RNN architecture with implementation justifications in
Section~\ref{sec:arch}. In Section~\ref{sec:evaluation} we
present the system evaluation followed by conclusions in Section~\ref{sec:conclusion}.
 
\section{Related work}\label{sec:related}

Yield forecast is an important service of agriculture computational
systems~\cite{holzworth2015agricultural,louhichi2010generic}. In this section we
cover a few efforts in this area, highlighting some with machine learning
components.

Kogan \emph{et al.}~\cite{kogan2013winter} compared different methods for winter
wheat yield forecasting: using remote sensing observations, meteorological data
and biophysical models. The two former methods consisted respectively of linear
regression models using NDVI data at 250m resolution and data from 180 weather
stations for a 13-year period as predictors. The third method is based on the
application of a biophysical process-based crop model, an algorithm that models
phenology, canopy development, biomass accumulation, water stress and many other
plant components. In this case, the World Food Studies (WOFOST)
model~\cite{boogaard1998wofost} was used. All three approaches were used to
perform forecast 2--3 months before harvest and the biophysical model showed the
best results in terms of root mean squared error (RMSE). The NDVI-based and the
meteorological data-based methods showed similar performance when minimum input
data requirements were met.

In studying dryland maize in South Africa, Estes \emph{et
al.}~\cite{estes2013comparing} developed three empirical models that were
compared against the CERES-maize model of the DSSAT
platform~\cite{jones2003dssat}. Two of the empirical models were based on
maximum entropy (MAXENT)~\cite{phillips2006maximum}: one trained on all national
crop distributions points and the other trained with the top producing
localities.  The third method used a generalized additive model (GAM) trained
with yield data derived from NDVI\@. GAM and CERES results showed linear
correlation to measured yield (R$^{2}$ = 0.75 and 0.37, respectively) as did the
MAXENT model trained with high-productivity points (R$^{2}$ = 0.62).

Gonzalez-Sanchez \emph{et al.}~\cite{gonzalez2014predictive} compared the
predictive accuracy of several techniques (Multiple linear regression, M5-Prime
regression trees, perceptron multilayer neural networks, support vector
regression and k-nearest-neighbors/KNN) for crop yield prediction in ten crop
datasets from western Mexico. Predictors came from typical atmospheric data
(solar radiation, rainfall, temperature, etc) and some genetic and farm
management information like season-duration cultivar and planting area. For
these specific conditions, the regression trees and KNN showed the lowest error
metrics.

Kumar~\cite{kumar2011crop} performed rice yield forecasting by adaptive neuro
fuzzy inference system (ANFIS) technique. For that, they used 27 years time
series data of yield and weather. ANFIS is an effort to integrate the benefits
of neural networks and fuzzy logic in a single framework by using linguistic
information from the fuzzy logic and learning capabilities of an artificial
neural network (ANN).  Quantitative performance assessment for rice and wheat
yield observations in India showed good applicability of the technique in yield
prediction.

Dahikar \emph{et al.}~\cite{dahikar2014agricultural} studied the basic
requirements for applications of ANNs in yield prediction. Simple network
architectures, with one hidden layer and back propagation of errors were tested
for different predictors and crops, like cotton, sugarcane, wheat, rice and
others. Soil parameters detected to be relevant for crop yield prediction were
PH and concentrations of nitrogen, phosphate, potassium, organic carbon,
calcium, magnesium, sulphur, manganese, copper and iron. In terms of atmospheric
predictors, temperature, rainfall and humidity were the relevant features
detected.

The proposed system differs from existing solutions as (i) it does not rely on NDVI
data, which is a more scalable approach to handle country-level forecasts, and
(ii) it is able to provide yield forecasts with a seven-month lead time, while offering similar
(or even better) results to systems that perform short term yield forecasts.
 
\section{Architecture}\label{sec:arch}

In this work, we combined atmospheric data (accumulated precipitation, maximum,
and minimum temperature) and soil data to produce a model capable of generating
yield forecast data---as illustrated in Figure~\ref{fig:system}. We implemented
a Deep Neural Network (DNN) as our machine learning model. This section provides
a description of the multiple data sources required to feed our DNN model, the
model itself, and its usage in production.

\begin{figure*}[!ht]
  \begin{center}
    \includegraphics[width=1.03\linewidth]{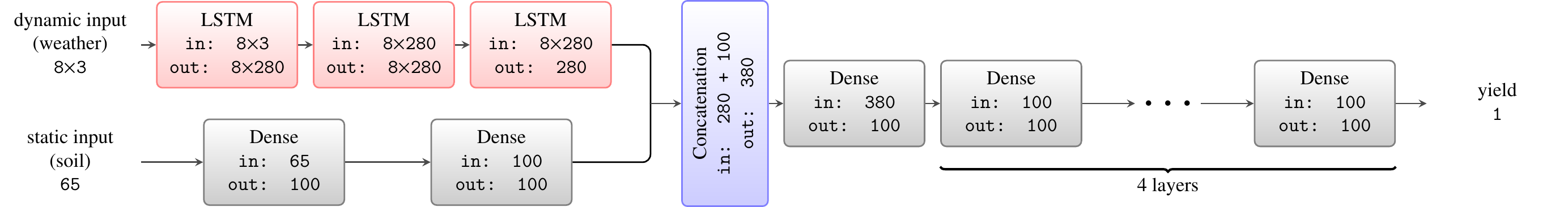}
  \end{center}
  \caption{%
      DNN Architecture. Red nodes represent Long Short-Rerm Memory (LSTM)
      recurrent layers. Gray nodes represent dense, fully-connected layers.
      The blue node represents a concatenation layer, which concatenates the
      intermediate representations from the dynamic and static paths. Numbers
      below node names represent shapes of input and output tensors. For
      example, \texttt{n$\times$m} represents a \texttt{n} by \texttt{m}
      matrix, while single numbers represent line vectors.
  }\label{fig:architecture}
\end{figure*}

\subsection{Data Sources}\label{sec:datasources}

Monthly precipitation data was provided by the Climate Hazards Group Infrared
Precipitation with Stations (CHIRPS) dataset~\cite{funk2015climate}. CHIRPS
provides precipitation data at 0.05$^{\circ}$ resolution by merging satellite
and weather station information. CHIRPS uses satellite data in three ways:
satellite means are used to produce high-resolution rainfall climatologies,
infrared Cold Cloud Duration (CCD) fields are used to estimate monthly and,
pentadal rainfall deviation from climatologies. Lastly, satellite precipitation
fields are used to guide interpolation through local distance decay functions.

Monthly air temperature data, specifically minimum and maximum temperatures for
each month, were provided by reanalysis datasets\footnote{Reanalysis datasets
are datasets produced both from observational data and numerical models.} from
ERA-Interim, one of the latest reanalysis products developed by the European
Centre for Medium-Range Weather Forecasts (ECMWF)~\cite{dee2011era}. This
dataset covers the globe at a resolution of approximately 80km per pixel and was
generated using data assimilation from several sources. The project represents
a significant improvement over past reanalysis efforts, due to advances in
modeling of several climate processes such as the hydrological cycle and
assimilation of cloud- and rain-affected satellite radiances.

Soil properties data comes from SoilGrids.org~\cite{hengl2017soilgrids250m},
an open, global soil dataset with a resolution of 250m per pixel which provides
information for clay, silt and sand contents plus fine earth and coarse
fragments bulk density. All this data is available in seven depths (0, 5, 15,
30, 60, 100 and 200cm). SoilGrids data are results of predictions based on
150000 soil profiles used for training and 158 remote sensing-based soil
covariates. These were used to fit an ensemble of random forest, gradient
boosting and multinomial logistic regression models.

As labels for training, the model uses actual yield data. In this work, we
collected maize and/or soybean yield data at county or municipality level from
each country's official agency. For Brazil, we used soybean data provided by
IBGE (Brazil Statistics and Geography Bureau)~\cite{ibgepma}, while for the USA
we used maize and soybean data provided by the USDA (United States Department
of Agriculture)~\cite{nass2018united}. Crop seasons in both countries last
approximately six months, although the tropical climate in Brazil allows for a
second mini-season in-between main seasons. The dataset sizes used in training,
validation, and testing are described in Table~\ref{tab:set_sizes}. For each
case, test sets were composed of 20\% of the total data points, while training
plus validation sets were composed of the remaining points (from which being
30\% validation and 70\% for training).

We split the input data into two main categories: \emph{static} for the soil
data and \emph{dynamic} for the weather data. The rationale behind this decision
is that for the time scales considered in this work soil properties do not change
over time, while meteorological data presents seasonal variability. Knowing the
data we used has different characteristics allows us to approach them
differently when building an ML model. The static features (corresponding to the
soil data) we used to train the model are shown in Table~\ref{tab:input_static},
and the dynamic features (corresponding to the atmospheric data) we used to
train the model are shown in Table~\ref{tab:input_dynamic}.

\begin{table}
\centering
\caption{%
    \emph{Static} features used in the Machine Learning model. All the soil features
    here are available in seven layers. Latitude and longitude of
    the prediction point are also included in the static inputs. Hence, there
    are $9\times7 + 2 = 65$ \emph{static} features.
}\label{tab:input_static}
\begin{tabular}{lr}
    \toprule
    Feature & Unit \\
    \midrule
    Clay content (0--2 micrometer) mass fraction & [\%]     \\
    Silt content (2--50 micrometer) mass fraction & [\%]    \\
    Sand content (50--2000 micrometer) mass fraction & [\%] \\
    Bulk density (fine earth) & [kg/m${3}$]                \\
    Coarse fragments volumetric & [\%]                     \\
    Cation exchange capacity of soil & cmolc/kg             \\
    Soil organic carbon content (fine earth fraction) & g per kg \\
    Soil pH x 10 in H2O & --- \\
    Soil pH x 10 in KCl & --- \\
    Point Latitude  &  \textdegree{} \\
    Point Longitude &  \textdegree{} \\
    \bottomrule
\end{tabular}
\end{table}

\subsection{Neural Network Description}

As we split the data into \emph{dynamic} and \emph{static} sets, the data follow
different pathways in the model before the joining the internal representations
of both data types (Figure~\ref{fig:architecture}). For the \emph{static} set, the data flows through a
two-layer fully-connected neural network before proceeding in the computational
graph---each layer with a hundred hidden units. The \emph{dynamic} set flows
through a three-layer LSTM neural network, each with 280 memory cells and eight
time steps (one for each month in the seasonal forecast), before proceeding in
the computational graph.  LSTM~\cite{hochreiter1997long} units are recurrent
neural network modules that are useful when dealing with data with a temporal
relationship, and can learn to recognize temporally extended patterns in noisy
sequences. Hence, they were chosen to model the \emph{dynamic} input in our
model.

\begin{table}
\centering
\caption{%
    \emph{Dynamic} Features used in the Machine Learning model. All the features
    in here are available for eight months covering the crop
    cycle in each geography (September to April in Brazil; April to October in
    US). Hence, there are $3\times8=24$ \emph{dynamic} features.
}\label{tab:input_dynamic}
\begin{tabular}{lr}
    \toprule
    Feature & Unit \\
    \midrule
    Minimum Temperature  & [°C]  \\
    Maximum Temperature  & [°C]  \\
    Precipitation        & [mm]  \\
    \bottomrule
\end{tabular}
\end{table}

The aforementioned describes the process outlined in the upper part of
Figure~\ref{fig:architecture} before the node labeled ``Concatenation Layer''.
When both data paths are computed, the network joins them together through the
concatenation of internal representations. After the concatenation, the network
processes the joined data through five fully-connected layers, each with
a hundred hidden units. Finally, the network outputs a single value: the
forecasted yield. The model uses the Mean Absolute Error as cost function and
uses scaled exponential linear units (SELUs)~\cite{klambauer2017self}. The SELU
activation function is given by
\begin{align}
    \mathrm{selu}(x) = \lambda
    \begin{cases}
        x                   & \text{if } x > 0,\\
        \alpha e^x - \alpha & \text{if } x \le 0
    \end{cases}\text{.}\label{eq:selu}
\end{align}
In Equation~\eqref{eq:selu}, $\alpha$ and $\lambda$ are chosen in a way that the mean and
variance of the inputs are preserved between two consecutive layers. Hence, such
an activation leads to Self-Normalizing Networks with the property of being
robust to perturbations, and not having high variance in training
errors~\cite{klambauer2017self}.

A full description of the model as implemented in Keras~\cite{chollet2015keras}
is shown in Figure~\ref{fig:source} and represents everything that is needed to
replicate the model described in this section, and ensures reproducibility of
the results described in this work (Section~\ref{sec:results}).

\begin{figure}[t]
\scriptsize
\tt
\noindent
\mbox{}dynamic$\_$input\ \textcolor{BrickRed}{=}\ \textbf{\textcolor{Black}{Input}}\textcolor{BrickRed}{(}shape\textcolor{BrickRed}{=(}\textcolor{Purple}{8}\textcolor{BrickRed}{,}\textcolor{Purple}{3}\textcolor{BrickRed}{),}\ dtype\textcolor{BrickRed}{=}\texttt{\textcolor{Red}{'float32'}}\textcolor{BrickRed}{)} \\
\mbox{}inner$\_$lstm1\ \textcolor{BrickRed}{=}\ \textbf{\textcolor{Black}{LSTM}}\textcolor{BrickRed}{(}\textcolor{Purple}{280}\textcolor{BrickRed}{,}\ return$\_$sequences\textcolor{BrickRed}{=}True\textcolor{BrickRed}{)(}dynamic$\_$input\textcolor{BrickRed}{)} \\
\mbox{}inner$\_$lstm2\ \textcolor{BrickRed}{=}\ \textbf{\textcolor{Black}{LSTM}}\textcolor{BrickRed}{(}\textcolor{Purple}{280}\textcolor{BrickRed}{,}\ return$\_$sequences\textcolor{BrickRed}{=}True\textcolor{BrickRed}{)(}inner$\_$lstm1\textcolor{BrickRed}{)} \\
\mbox{}lstm$\_$out\ \textcolor{BrickRed}{=}\ \textbf{\textcolor{Black}{LSTM}}\textcolor{BrickRed}{(}\textcolor{Purple}{280}\textcolor{BrickRed}{)(}inner$\_$lstm2\textcolor{BrickRed}{)} \\
\mbox{} \\
\mbox{}static$\_$input\ \textcolor{BrickRed}{=}\ \textbf{\textcolor{Black}{Input}}\textcolor{BrickRed}{(}shape\textcolor{BrickRed}{=(}\textbf{\textcolor{Black}{len}}\textcolor{BrickRed}{(}stat$\_$cols\textcolor{BrickRed}{),))} \\
\mbox{}inner$\_$stat1\ \textcolor{BrickRed}{=}\ \textbf{\textcolor{Black}{Dense}}\textcolor{BrickRed}{(}\textcolor{Purple}{100}\textcolor{BrickRed}{,}\ activation\textcolor{BrickRed}{=}\texttt{\textcolor{Red}{'selu'}}\textcolor{BrickRed}{)(}static$\_$input\textcolor{BrickRed}{)} \\
\mbox{}inner$\_$stat2\ \textcolor{BrickRed}{=}\ \textbf{\textcolor{Black}{Dense}}\textcolor{BrickRed}{(}\textcolor{Purple}{100}\textcolor{BrickRed}{,}\ activation\textcolor{BrickRed}{=}\texttt{\textcolor{Red}{'selu'}}\textcolor{BrickRed}{)(}inner$\_$stat1\textcolor{BrickRed}{)} \\
\mbox{} \\
\mbox{}x\ \textcolor{BrickRed}{=}\ \textbf{\textcolor{Black}{concatenate}}\textcolor{BrickRed}{([}lstm$\_$out\textcolor{BrickRed}{,}\ inner$\_$stat2\textcolor{BrickRed}{])} \\
\mbox{}\textbf{\textcolor{Blue}{for}}\ $\_$\ \textbf{\textcolor{Blue}{in}}\ \textbf{\textcolor{Black}{range}}\textcolor{BrickRed}{(}\textcolor{Purple}{5}\textcolor{BrickRed}{):} \\
\mbox{}\ \ \ \ x\ \textcolor{BrickRed}{=}\ \textbf{\textcolor{Black}{Dense}}\textcolor{BrickRed}{(}\textcolor{Purple}{100}\textcolor{BrickRed}{,}\ activation\textcolor{BrickRed}{=}\texttt{\textcolor{Red}{'selu'}}\textcolor{BrickRed}{)(}x\textcolor{BrickRed}{)} \\
\mbox{} \\
\mbox{}dynamic$\_$output\ \textcolor{BrickRed}{=}\ \textbf{\textcolor{Black}{Dense}}\textcolor{BrickRed}{(}\textcolor{Purple}{1}\textcolor{BrickRed}{,}\ activation\textcolor{BrickRed}{=}\texttt{\textcolor{Red}{'selu'}}\textcolor{BrickRed}{)(}x\textcolor{BrickRed}{)} \\
\mbox{} \\
\mbox{}model\ \textcolor{BrickRed}{=}\ \textbf{\textcolor{Black}{Model}}\textcolor{BrickRed}{(}inputs\textcolor{BrickRed}{=[}dynamic$\_$input\textcolor{BrickRed}{,}\ static$\_$input\textcolor{BrickRed}{],} \\
\mbox{}\ \ \ \ \ \ \ \ \ \ \ \ \ \ outputs\textcolor{BrickRed}{=[}dynamic$\_$output\textcolor{BrickRed}{])} \\
\mbox{}model\textcolor{BrickRed}{.}\textbf{\textcolor{Black}{compile}}\textcolor{BrickRed}{(}loss\textcolor{BrickRed}{=}\texttt{\textcolor{Red}{'mae'}}\textcolor{BrickRed}{,} \\
\mbox{}\ \ \ \ \ \ \ \ \ \ \ \ \ \ optimizer\textcolor{BrickRed}{=}\textbf{\textcolor{Black}{Adam}}\textcolor{BrickRed}{(}lr\textcolor{BrickRed}{=}\textcolor{Purple}{0.0005}\textcolor{BrickRed}{))} \\
\mbox{}
   \caption{%
      DNN used in this work as a model implemented in Keras.
  }\label{fig:source}
\end{figure}

\subsection{Production System}

\begin{figure*}[ht!]
  \centering
  \begin{subfigure}[b]{0.48\linewidth}
    \includegraphics[width=\linewidth]{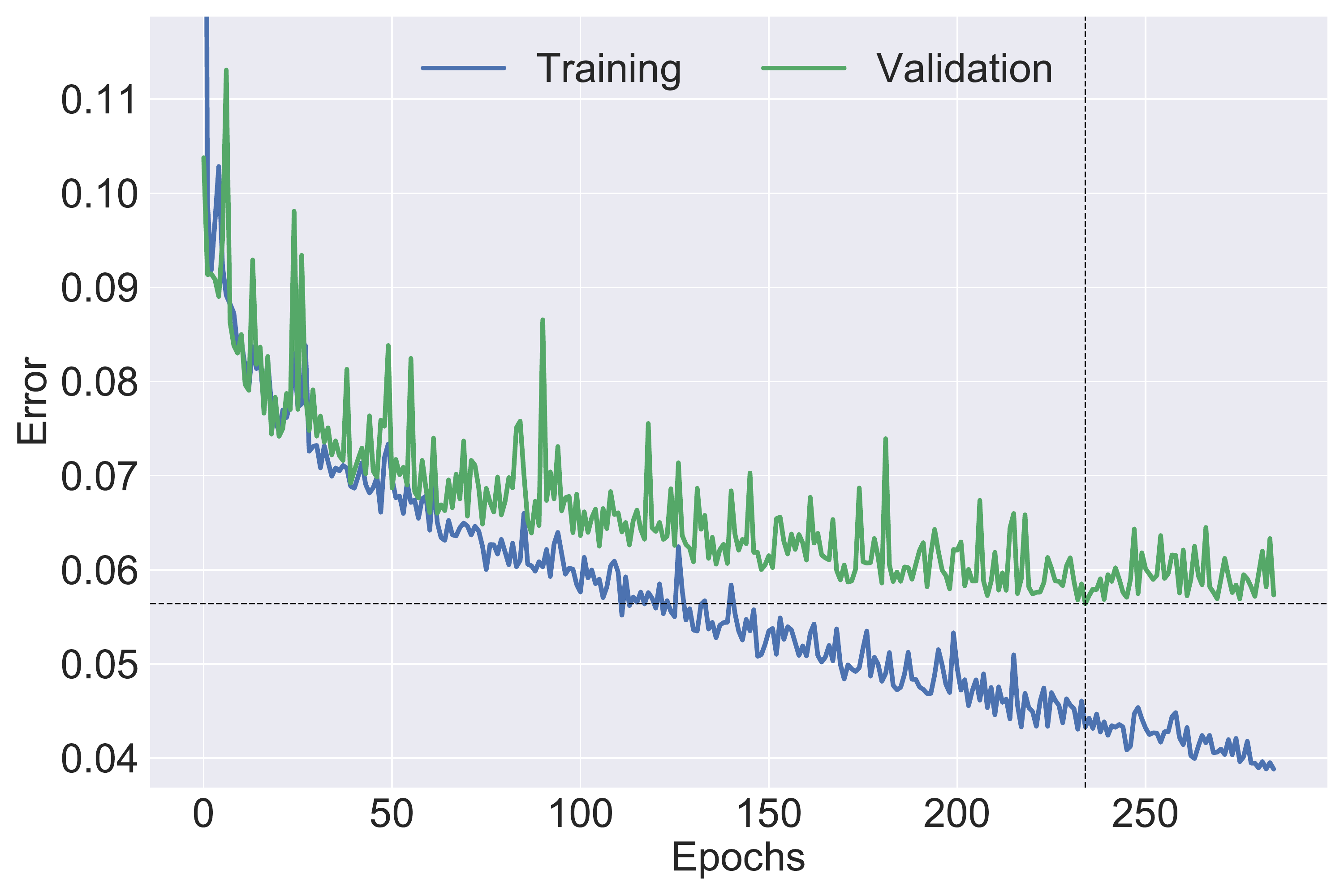}
     \caption{US soybean model}\label{fig:training_a}
  \end{subfigure}
  \begin{subfigure}[b]{0.48\linewidth}
    \includegraphics[width=\linewidth]{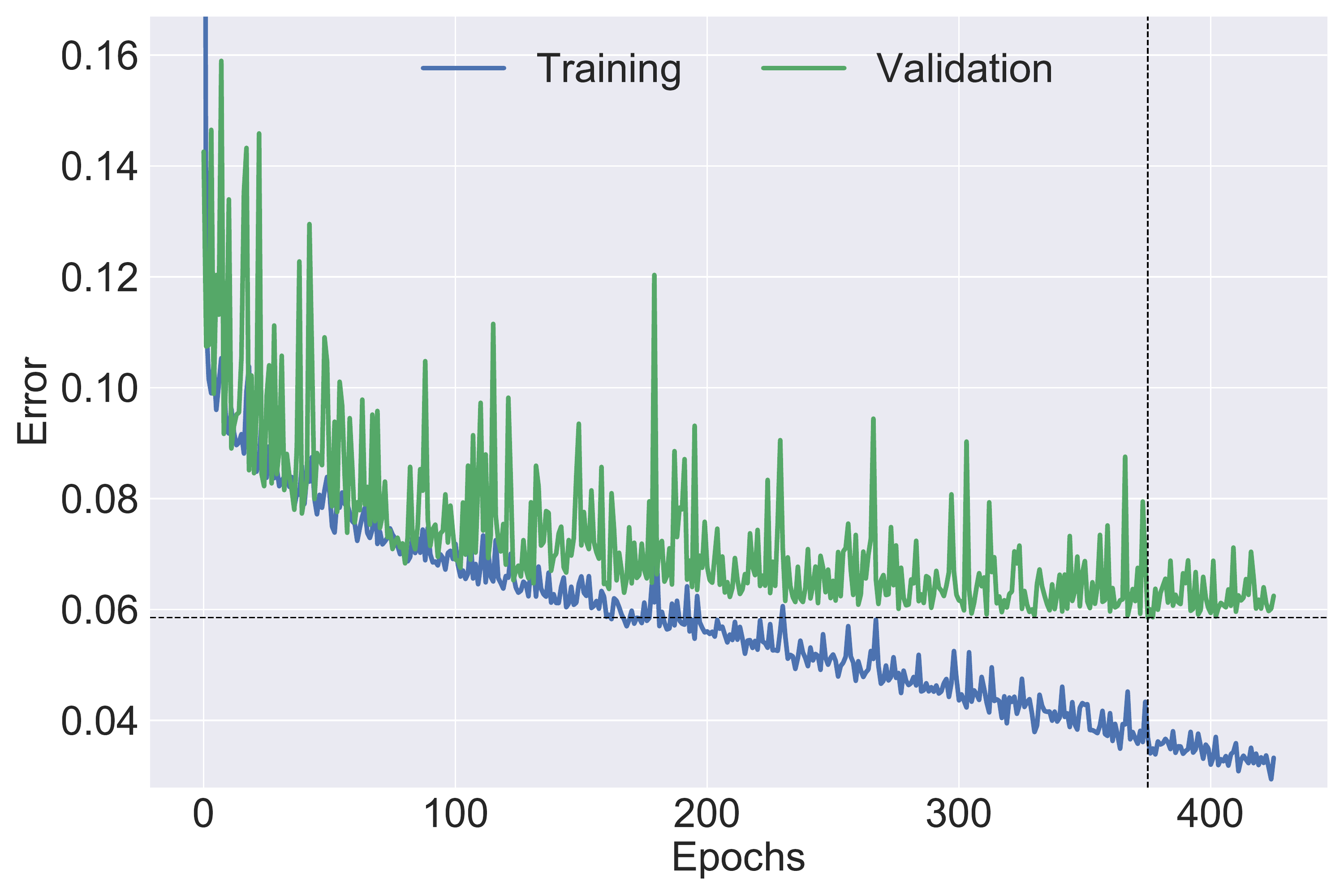}
    \caption{US maize model}\label{fig:training_b}
  \end{subfigure}
  \begin{subfigure}[b]{0.48\linewidth}
    \includegraphics[width=\linewidth]{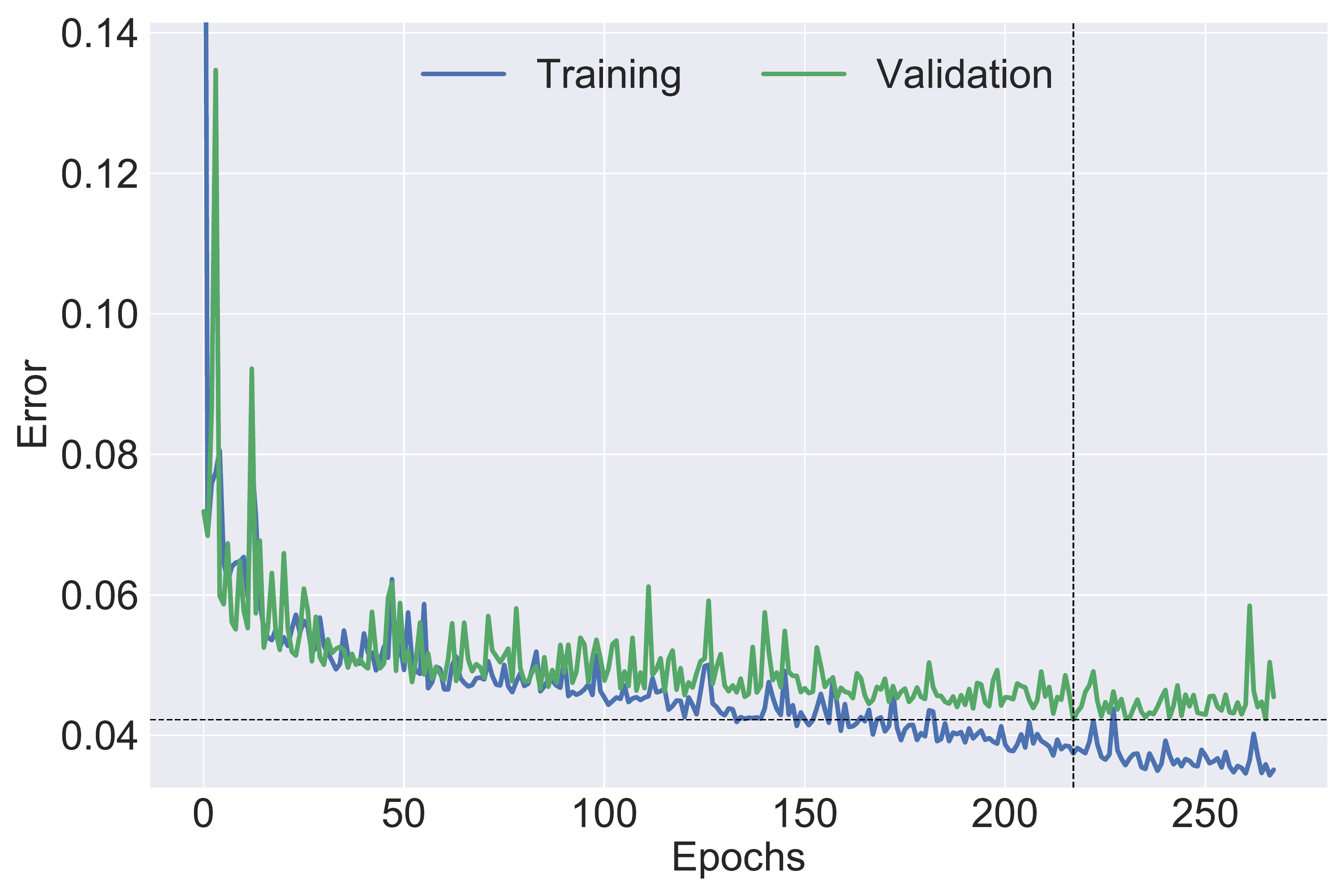}
    \caption{Brazil soybean model}\label{fig:training_c}
  \end{subfigure}
  \caption{
      Loss for training and validation sets across epochs
      for Brazil/USA soybean/maize models. Dashed lines indicate
      the chosen model, which is the one with minimum validation loss.
  }\label{fig:training}
\end{figure*}

As described in the previous section, after the DNN is fully trained, one can
forecast the yield of a single point: given a pair $p
= (\mathrm{latitude}\text{, }\mathrm{longitude})$, one can query the
meteorological and soil properties datasets, extract the data corresponding to
$p$ and perform a forecast. Although this would work for any point in the globe,
the model was trained with Brazil and US yield data so it wouldn't make
practical sense to use it outside these geographies. Moreover, it is possible
to integrate the DNN into a bigger system for on-the-fly yield forecast. We
implemented the DNN as a component in a decision-support tool for agriculture.
To forecast yields for future dates, we replaced the ERA-Interim data with
seasonal forecasts (which are also based on reanalysis data) for the
\emph{dynamic} features and continued using SoilGrids data. Due to its
relatively high resolution, the SoilGrids data allows the system to also
forecast with a resolution of 250m per pixel.

\begin{figure*}[ht!]
  \centering
  \begin{subfigure}[b]{0.48\linewidth}
    \includegraphics[width=\linewidth]{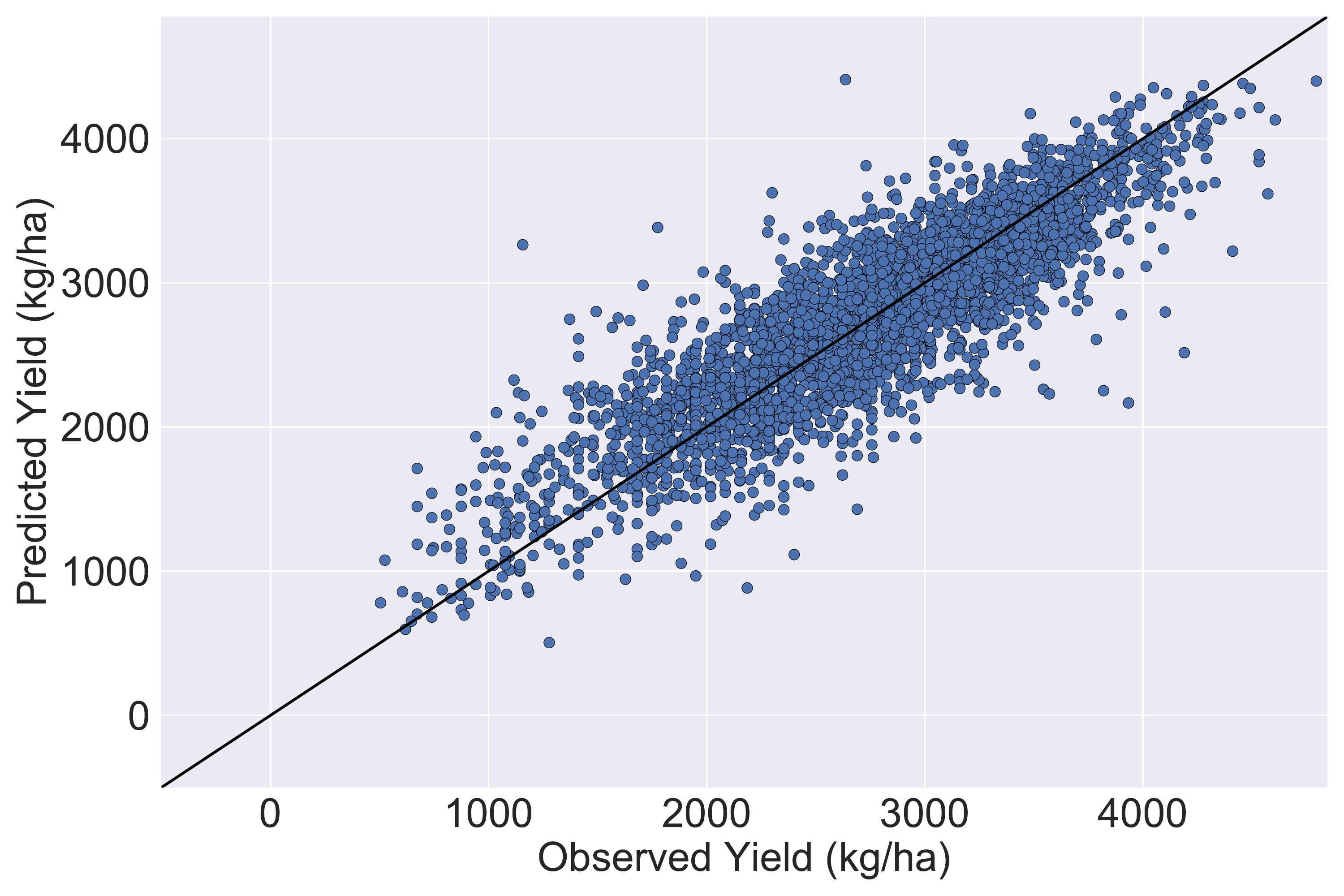}
     \caption{US soybean model}\label{fig:scatter_a}
  \end{subfigure}
  \begin{subfigure}[b]{0.48\linewidth}
    \includegraphics[width=\linewidth]{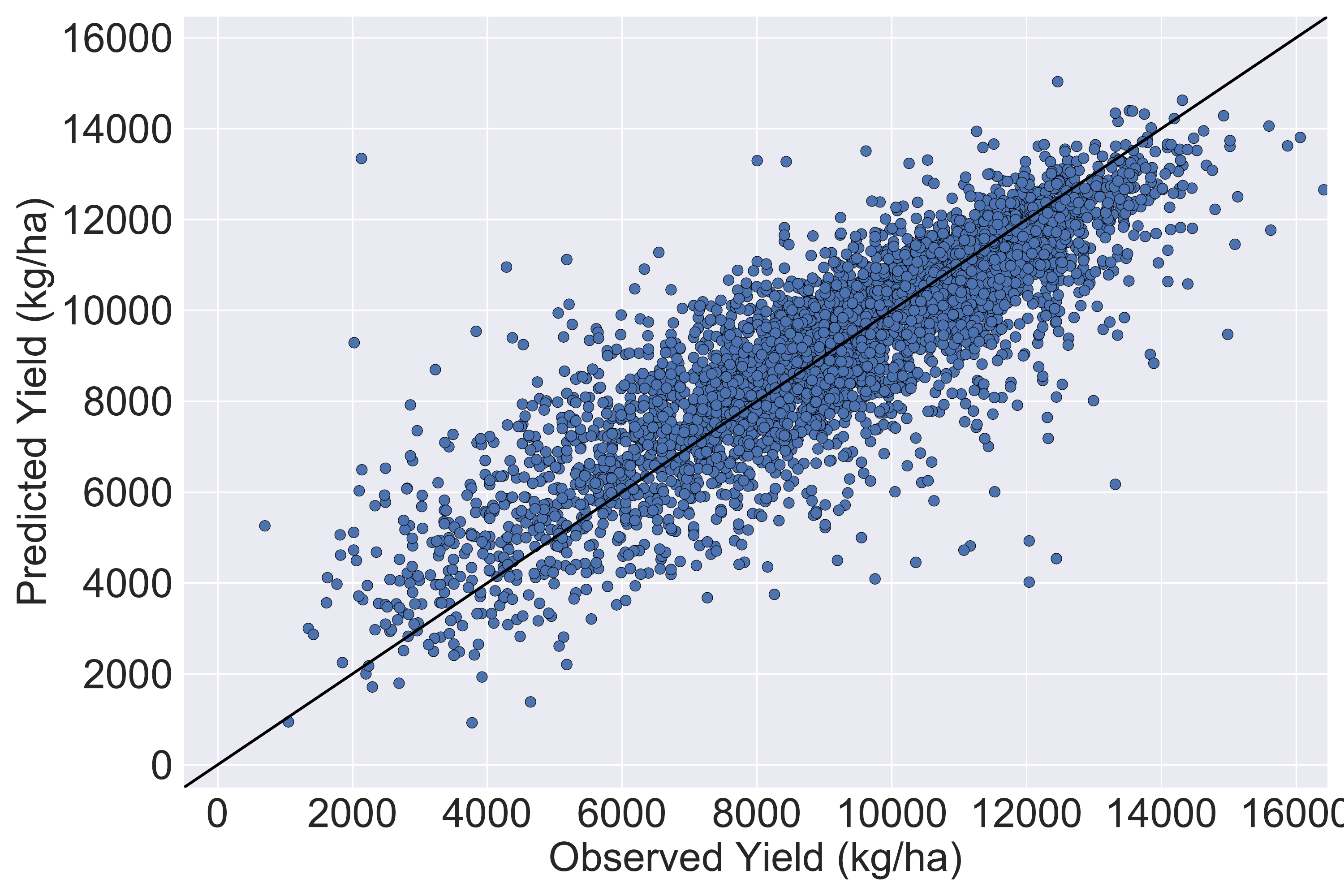}
    \caption{US maize model}\label{fig:scatter_b}
  \end{subfigure}
  \begin{subfigure}[b]{0.48\linewidth}
    \includegraphics[width=\linewidth]{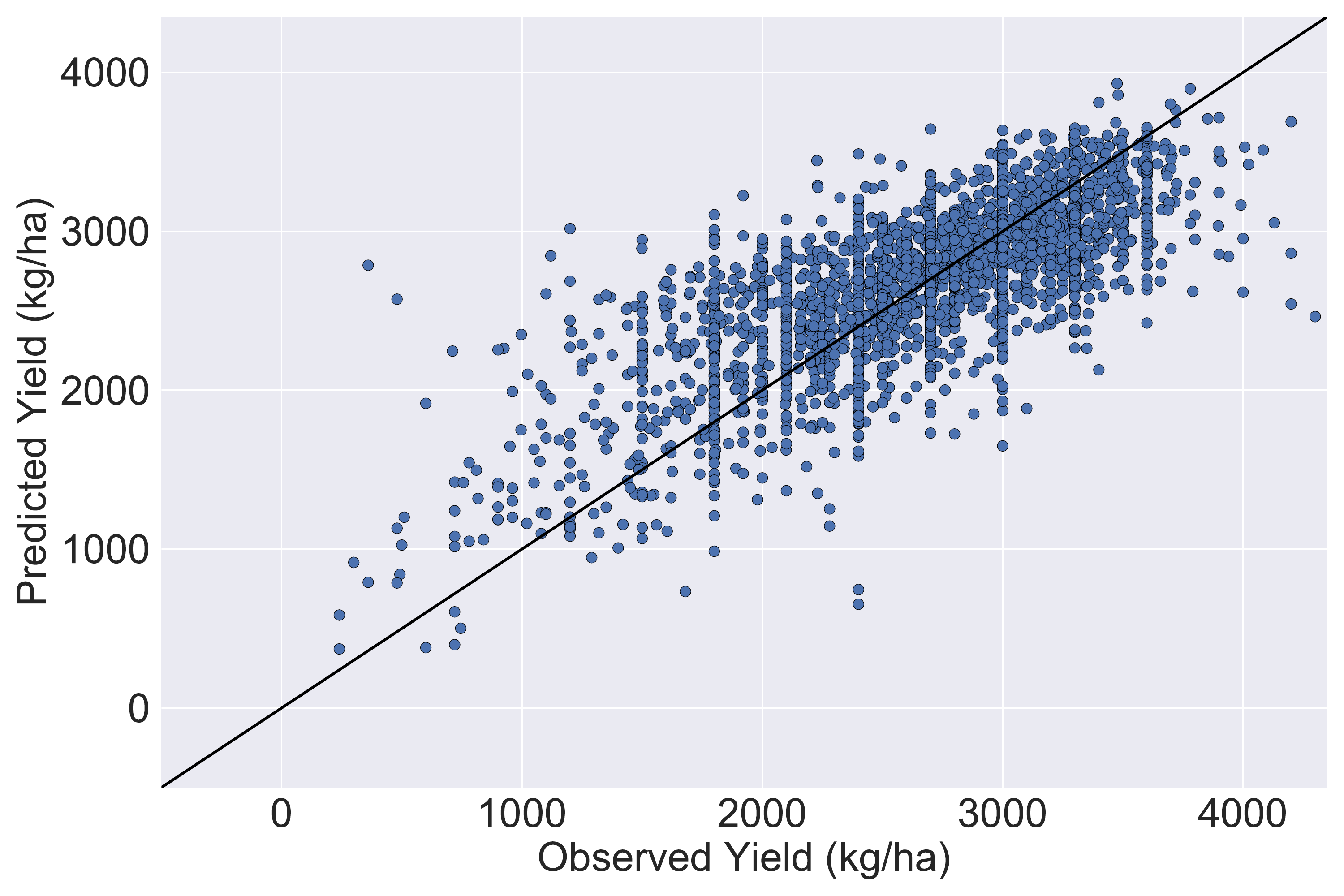}
    \caption{Brazil soybean model}\label{fig:scatter_c}
  \end{subfigure}
  \caption{
      Scatter plots of model performance according to country and crop.
  }\label{fig:scatter}
\end{figure*}
 
\section{Evaluation}\label{sec:evaluation}

The goal of the evaluation is to demonstrate the effectiveness of the proposed
yield forecast model against existing solutions that rely on remote sensing
data, \textit{e.g.} NDVI\@. We used five comparison metrics in the evaluation
related to accuracy of the yield forecast.

\subsection{Experimental Setup}

Data for the 3 cases (US-Soybean, US-Maize and Brazil-Soybean)
was used in training and testing the DNN model. For each case, 80\% of the total
data points were selected as training set and the remaining 20\% as test set. During
training, 30\% of the training set was selected for a validation set. All of these sets
were randomly chosen, so the model could learn from many different scenarios of weather,
soil and forms of growing. The sizes of each set are detailed in Table~\ref{tab:set_sizes}.

Model training for each case took around 300 epochs until
a minimum validation loss value was achieved. The evolution of training and validation
losses is shown in Figure~\ref{fig:training}. Losses show a typical neural
network behaviour of continuous decay for training loss, eventually decoupling from validation
loss. The training made use of the \texttt{EarlyStopping} callback function from the keras
library, with a \texttt{patience} parameter (the number of epochs with no improvement after
which training is stopped) equal to 50. The DNN model could be successfully trained using a relatively
modest GPU (Tesla K40m) in reasonable time (\texttildelow10 minutes). Test sets for each
geography and crop example were used for evaluation of the model performance in forecasting
crop yield at pre-season time and are discussed in Section~\ref{sec:results}.

\begin{table}
\centering
\caption{%
    Datasets sizes used in training, testing and validation.
}\label{tab:set_sizes}
\begin{tabular}{lrrr}
\toprule
             & Brazil-Soybean &  US-Soybean & US-Maize  \\
\midrule
\# of counties        &  1529 &  1814 &  2204  \\
\# of data points     & 16767 & 16939 & 19692  \\
Training set size     &  9389 &  9485 & 11027  \\
Validation set size   &  4023 &  4065 &  4725  \\
Test set size         &  3354 &  3388 &  3819  \\
\bottomrule
\end{tabular}
\end{table}
 
\subsection{Result Analysis}\label{sec:results}

The model was used to perform yield forecast for the points in the test set and the results
shown in Table~\ref{tab:scores}. Since the training and test sets were randomly selected
from the complete input dataset, a good performance in the test set indicates the model
generalizes well for different seasonal climates, agriculture management practices, soil and
other geographical characteristics. Figure~\ref{fig:scatter} shows comparison of observed and
forecasted yield for each geography and crop.

When comparing geographies, model results for the US are slightly better than the ones found for
Brazil. It is important to notice the specific methodologies in gathering
yield data for both countries. The former relies more on self-reporting information provided by
farmers~\cite{ibgepma}, whereas in the latter case there is a data gathering  effort
by the governmental agency~\cite{nass2018united}. These different methods are reflected on both raw
datasets: the one for Brazil showed a higher number of missing information and in a relevant number
of counties the reported yield remains unchanged along some years, which suggest the data is a
gross estimation.

Model performance of the US-Soybean forecasts was superior to US-Maize. This initially indicates
that the DNN architecture is better in creating a function that maps the physical relation
between soil plus climate to yield in the soybean case. Physiological differences between soybean
and maize determine the performance in each case. To account for the temperature influence on both
crops, we can analyze the typical values of accumulated growing degree days (AGDD). Soybean
typically has lower AGDD values (around 1100 °C)~\cite{akyuz2017developing} than Maize (around
1450 °C)~\cite{martin2017determining}. In terms of water needs, studies have showed maize has a
higher vulnerability to moisture deficiency when compared to soybean~\cite{zipper2016drought}.
All of these indicate that maize is more sensitive to the climate variables (temperature and
precipitation) used in model training than soybean, meaning any uncertainties brought by the
climate data input sources will have a stronger impact in the maize yield forecast, which could
explain the difference in performance of the model for the two crops.

The comparison of these results to other yield forecasting efforts should take in consideration
several differences amongst these studies, including the selected statistics reported and its
appropriate units. Metrics like the coefficient of determination $R{^2}$ and the ones that
compute errors as a ratio of some average field (MAPE, RMSPE) can be compared across different
crops and regions, besides being commonly reported. Table~\ref{tab:comparison} shows some of these
metrics reported in several studies. All of these efforts make use of some sort of remote-sensing,
most frequently NDVI\@. The performance of this work is comparable with these studies but with a lot
less data requirements and is able to provide useful information for agriculture operations before
the seeding occurs.

\begin{table}
\centering
\caption{Model scores.}\label{tab:scores}
\begin{tabular}{lrrr}
\toprule
             & Brazil - Soybean &  US - Soybean & US - Maize  \\
\midrule
MAE$^a$      &  288.39 &  270.18 &  1031.00  \\
MAPE$^b$     & 10.70\% &  9.80\% &  11.31\%   \\
RMSE$^c$     &  385.81 &  354.08 &  1393.02   \\
RMSPE$^d$    & 14.31\% & 12.85\% &  15.28\%   \\
R${^2}$$^e$ &    0.55 &     0.75 &     0.71   \\
\bottomrule
\multicolumn{4}{l}\footnotesize{$^a$ Mean Absolute Error}\\
\multicolumn{4}{l}\footnotesize{$^b$ Mean Absolute Percentage Error}\\
\multicolumn{4}{l}\footnotesize{$^c$ Root Mean Square Error}\\
\multicolumn{4}{l}\footnotesize{$^d$ Root Mean Square Percentage Error}\\
\multicolumn{4}{l}\footnotesize{$^e$ Coefficient of determination}\\
\end{tabular}
\end{table}

\begin{table*}
\centering
\caption{Comparison of yield forecasting methods to the ML system proposed in this work.}
\label{tab:comparison}
\begin{tabular}{rrrrrr}
\toprule
Study                                                         & Crop            & R${^2}_{\alpha}$   & MAPE$_{\beta}$ [\%]     & RMSPE$_{\gamma}$[\%]    &  ML System\\
\midrule
Kolotii, A \textit{et al.}, 2015~\cite{kolotii2015comparison}  & Wheat                 &  0.50 - 0.80 &  -            &  -           &  0.55 - 0.75$_{\alpha}$\\
Capa-Morocho, M \textit{et al.}, 2016~\cite{capa2016crop}      & Wheat, Maize          &  -           &  -            &  2.1 - 13.2  & NA - 15.28$_{\gamma}$\\
Meroni, M \textit{et al.}, 2016~\cite{meroni2016evaluating}    & Grain crops           &  0.62 - 0.91 &  -            &  -           & 0.55 - 0.75$_{\alpha}$\\
Morell, F \textit{et al.}, 2016~\cite{morell2016can}           & Maize                 &  -           &  -            &  20 - 34     & 2.85 - 15.28$_{\gamma}$\\
Johnson, D 2014~\cite{johnson2014assessment}                  & Maize, Soybean        &  0.47 - 0.77 &  -            &  -           & 0.55 - 0.75$_{\alpha}$\\
Pagani, V \textit{et al.}, 2018~\cite{pagani2018high}          & Rice                  &  0.21 - 0.89 &  -            &  -           & 0.55 - 0.75$_{\alpha}$\\
Kumar, N \textit{et al.}, 2014~\cite{kumar2014crop}            & Rice, Wheat           &  0.53 - 0.58 &  -            &  -           & 0.55 - 0.75$_{\alpha}$\\
Bose, P \textit{et al.}, 2016~\cite{bose2016spiking}           & Wheat                 &  -           &  0.13 - 27.97 &  -           & 9.8 - 11.31$_{\beta}$\\
Satir, O and Berberoglu, S, 2016~\cite{satir2016crop}         & Maize, Wheat, Cotton  &  0.50 - 0.67 &  6.30 - 7.30  &  -           & 0.55 - 0.75$_{\alpha}$ \& 9.8 - 11.31$_{\beta}$\\
\bottomrule
\multicolumn{6}{l}\footnotesize{${\alpha}$, ${\beta}$, ${\gamma}$ subscripts indicate which metric is being shown for the ML System}\\
\end{tabular}
\end{table*}
 
\section{Conclusion}
\label{sec:conclusion}

Agriculture yield forecasts are a very useful tool for farm management and can help stakeholders
to perform critical decisions in their agricultural operations. Many available methods provide
yield forecast information, the vast majority of them through the use of some sort of remote
sensing data (like NDVI) from the farm fields. While this allows for high-resolution forecasts, it
comes with the high cost of acquiring and processing these extra datasets, which can be relevant
depending on the properties dimensions.

Machine-learning is one of the techniques gaining popularity for agriculture applications,
specially with the increasing number of new data sources being developed in the latest years. We
propose a machine learning system that provides pre-season yield forecasting, meaning farmers can
make farm management decisions (like choosing different crops or genetic variations) before seeding
occurs.

The system proposed in this work is formed by a neural network where inputs are
treated separately. Static soil data in handled by fully-connected layers while
dynamic meteorological data is handled by recurrent LSTM layers. This particular
architecture was trained with historical data for several soil properties,
precipitation, minimum and maximum temperature against historical yield labels
at county level for two crops (maize and soybean) and two geographies (Brazil
and US), which production correspond to 44\% of the global grain
production~\cite{faostat2016}. After training, the model was tested in
a separate dataset and showed comparable results with existing yield forecasting
methods that make use of extensive remote-sensing data. The major lesson learnt
from our experiments is that it is possible obtain scalable yield forecast
because the proposed neural network model can detect and exploit redundant
information both in the soil and in the weather data. Additionally, the model
may have been able to learn an implicit representation of the cycles of the
crops evaluated in this paper, considering the seasonal atmospheric data used as
input.

Our results show that farmers and agriculture stakeholders can benefit from
useful information with significantly fewer data requirements and maintain useful
accuracy values. The global availability of the input datasets also allows the
system to easily scale across different regions if local yield data is present.
Although the used input datasets allow for relatively high-resolution forecasts
(250m), the existing system can be used as a foundation for future precision
agriculture services by assimilating more traditional NDVI and similar datasets.

\bibliographystyle{IEEEtran}
\balance
\bibliography{ref}
\end{document}